\RequirePackage{lineno} 
\documentclass[article]{revtex4}
\usepackage{graphicx}
\usepackage{dcolumn}
\usepackage{bm}
\usepackage{color}
\usepackage{eso-pic}
\usepackage{amsmath}
\usepackage{xspace}

\newcommand{\STO}{$\mathrm{SrTiO}_3$\xspace}
\newcommand{\LAO}{$\mathrm{LaAlO}_3$\xspace}
\newcommand{\TiO}{$\mathrm{TiO}_2$\xspace}

\newcommand{\LXO}{$\mathrm{LaXO}_3\mathrm{(X=Al,Ti)}$\xspace}
\begin{document}

\title{Quantized conductance in a one-dimensional ballistic oxide nanodevice}

\author{A. Jouan$^{1}$, G. Singh$^{1}$, E. Lesne$^{2}$, D. C. Vaz$^{2}$, M. Bibes$^{2}$, A. Barth\'el\'emy$^{2}$, C. Ulysse$^{3}$, D. Stornaiuolo$^{4,5}$, M. Salluzzo$^{5}$, S. Hurand$^{1,6}$, J. Lesueur$^{1}$, C. Feuillet-Palma$^{1}$, N. Bergeal$^{1}$}

\affiliation{$^{1}$Laboratoire de Physique et d'Etude des Mat\'eriaux - ESPCI-Paris-CNRS-UPMC\\PSL Research University 10 rue Vauquelin 75005 Paris, France.\\
$^{2}$Unit\'e Mixte de Physique CNRS-Thales,Univ. Paris-Sud, Universit\'e Paris-Saclay, 1 Av. A. Fresnel, 91767 Palaiseau, France.\\
$^{3}$Centre for Nanoscience and Nanotechnology, CNRS, Universit\'e Paris-Sud/Universit\'e Paris-Saclay, Boulevard Thomas Gobert, Palaiseau, France.\\
$^{4}$Department of Physics, University of Naples Federico II Complesso Monte S'Angelo via Cithia, 80126 Napoli, Italy.\\ 
$^{5}$CNR-SPIN Complesso Monte S. Angelo via Cinthia 80126 Napoli, Italy.\\
$^{6}$Institut Pprime, UPR 3346 CNRS, Universit\'e de Poitiers, ISAE-ENSMA, BP 30179, 86962 Futuroscope-Chasseneuil Cedex, France.}

\date{\today}
\large

\maketitle

\large 
\textbf{Electric-field effect control of two-dimensional electron gases (2-DEG) has enabled the exploration of nanoscale electron quantum transport in semiconductors.  Beyond these classical materials, transition metal-oxide-based structures have $d$-electronic states favoring the emergence of novel quantum orders absent in conventional semiconductors. In this context, the \LAO/\STO interface that combines gate-tunable superconductivity and sizeable spin-orbit coupling is emerging as a promising platform to realize topological superconductivity. However, the fabrication of nanodevices in which the electronic properties of this oxide interface can be controlled at the nanoscale by field-effect remains a scientific and technological challenge. Here, we demonstrate the quantization of conductance in a ballistic quantum point contact (QPC), formed by electrostatic confinement of the \LAO/\STO 2-DEG with a split-gate. Through finite source-drain voltage, we perform a comprehensive spectroscopic investigation of the $3d$ energy levels inside the QPC, which can be regarded as a spectrometer able to probe Majorana states in an oxide 2-DEG.}\\

 The interplay between superconductivity and spin-orbit coupling (SOC) is at the center of an intensive research effort since this unique combination can promote topological superconductivity, an exotic electronic state with remarkable chiral properties \cite{alicea}.  In particular, topological superconductors are predicted to be suitable hosts for Majorana zero energy modes, which could be used to encode and manipulate non-local quantum information, opening new perspectives for the realization of "fault tolerant" quantum computation technology \cite{stern,dassarma}. While most of experiments in this field focus on hybrid devices made of a semiconductor nanowires coupled to conventional superconductors \cite{mourik, finck,churchill,zhang}, oxide heterostructures offer a new material-based approach to this question.  Transition metal oxides present a wide variety of electronic properties, from superconductivity to magnetism that, combined in artificial heterostructures with atomically sharp interface, provide a unique opportunity for the generation of novel emergent properties and functionalities \cite{hwang}. The \LAO/\STO interface, discovered by Ohtomo and Hwang in 2004, is a prime example \cite{Ohtomo:2004p442}. While both materials are wide-gap band insulators, growing a thin layer of  \LAO on top of \STO leads to the formation of a high mobility 2-DEG at the interface, whose carrier density can be tuned by means of top-, side- and  back-gating  \cite{hurand,stornaiuoloPRB, Caviglia:2008p116}. Beyond a certain filling threshold,  likely related to a Lifshitz transition in the band structure \cite{valentinis,singh}, the interface exhibits superconductivity with a gate-tunable critical temperature. The breaking of translational symmetry in the interfacial quantum well also generates a spin-orbit coupling of Rashba type, whose strength can be tuned by gate voltage \cite{caviglia2,benshalom,singhCR,lesne}. Superconductivity and Rashba spin-orbit coupling, which remarkably coexist in \LAO/\STO, are the two main ingredients required to generate superconducting topological phases \cite{alicea,fu,stornaiuoloPRB2017, kuerten}. The \LAO/\STO interface could therefore serve as platform to fabricate devices in which these two properties could be tuned at the relevant scales using a set of nano-gates. In this article, we report on the fabrication of a QPC at a \LAO/\STO interface, a prototypical mesoscopic device made by the deposition of a metallic split-gate on a  2-DEG \cite{vanvwees}.  The observation of well-defined plateaus in the conductance of the device, quantized in integer values of $2e^2/h$ (where $e$ is the electron charge and $h$ is the Planck constant), indicates ballistic transport involving a limited number of conduction channels which are spin degenerate.  By measuring non-linear transport at finite source-drain voltage, we perform a spectroscopy of the $3d$-levels and extract the characteristic energies of the confinement potential. Under a magnetic field, we observe the formation of spin-polarized subbands and determine the Land\'e $g$-factor,  whose value of order of 1 strongly differs from that of the free electrons. The demonstration of quantized conductance in \LAO/\STO QPC is the first step towards the realization of a spectrometer able to provide distinctive signatures of topological superconductivity in oxide 2-DEGs \cite{wimmer,beenakker}. \\

\begin{figure}[b]
  \centering
  \includegraphics[width=12cm]{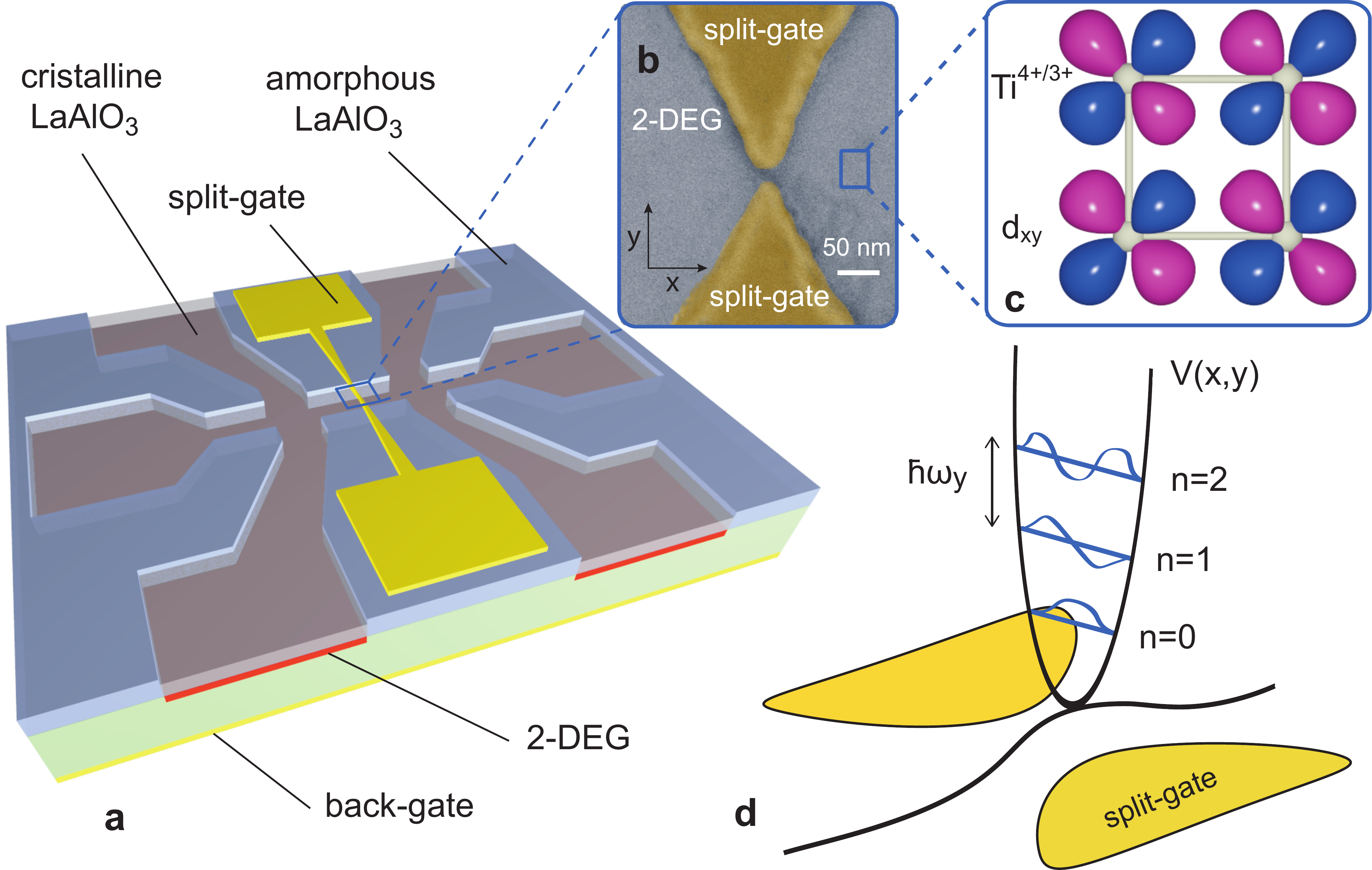}
  \caption{\textbf{Schematic description of the Quantum Point Contact device.} a) Hall Bar designed by the amorphous \LAO technique and covered with a split-gate. b) SEM image of the QPC showing the central part of the split-gate (false color). The separation between the two sides of the split-gate is  25 nm, c) Scheme of the $3d_\mathrm{xy}$ orbitals localised on Ti ions, which form the lowest energy conducting band in the 2-DEG. d) Representation of the confining potential at the center of the split-gate described by Eq. (\ref{potential}). Within  the harmonic approximation, the lowest $3d_\mathrm{xy}$ band is split into several sub-levels ($n$ = 1,2,3..) separated by an energy $\hbar\omega_y$.}
  \label{fig1}
\end{figure}

\begin{figure}[h!]
  \centering
  \includegraphics[width=8cm]{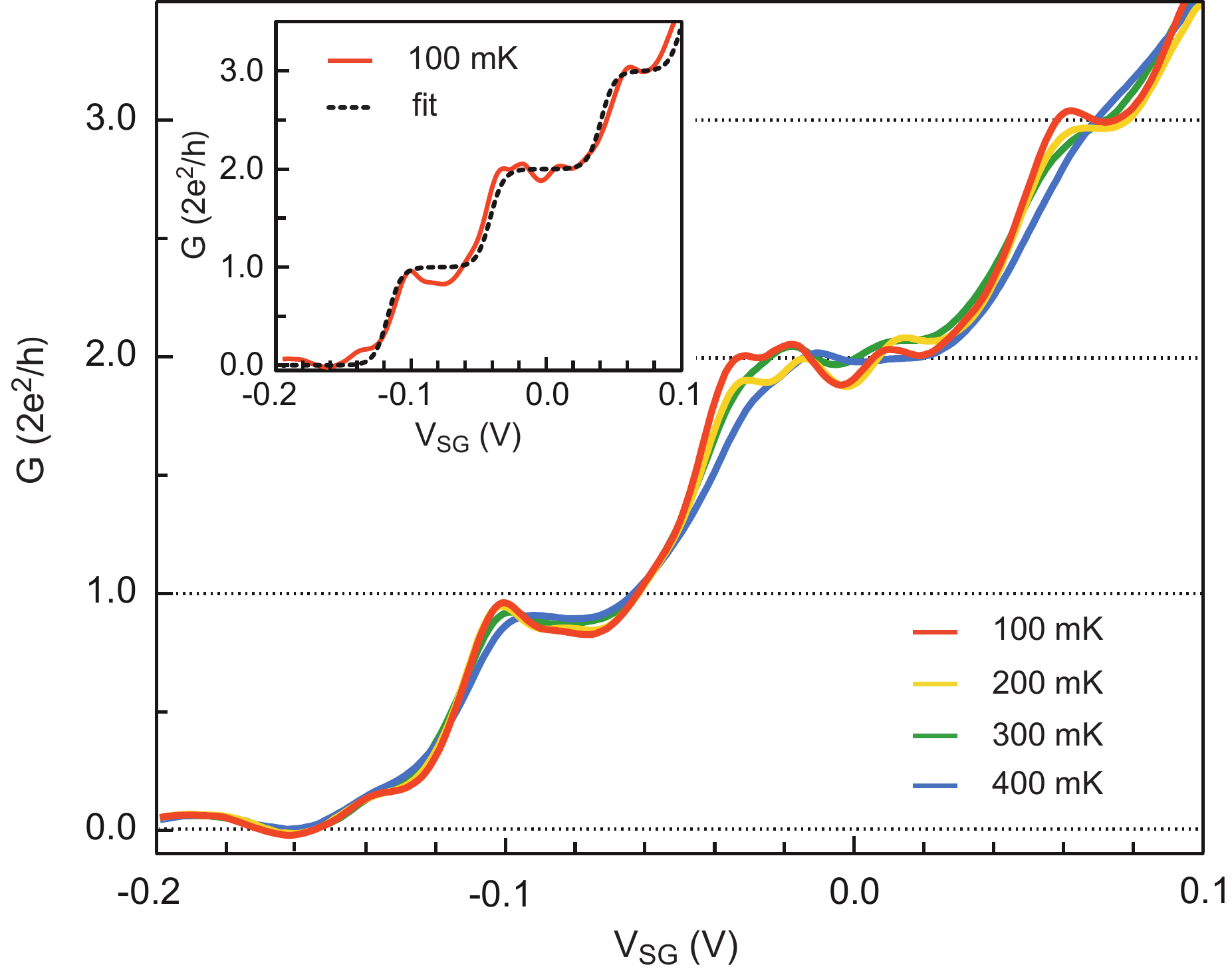}
  \caption{\textbf{Ballistic transport and conductance quantization}. Differential conductance of the QPC  as a function of split-gate voltage  measured at $V_{sd}=0$ for different temperatures and a fixed back-gate voltage $V_\mathrm{BG}$ = 8.1 V. Plateaus appear at integer value of $G_0$ characteristic of ballistic transport involving a reduced number of conduction modes. Inset) Fit of the $G(V_\mathrm{SG})$ curve at T = 100 mK (dashed line) obtained with  Eq. (2) by taking a ratio $\omega_y/\omega_x$ $\simeq$ 2.5.}
  \label{fig2}
\end{figure}

In a QPC, quantization of conductance is observed when the width of the constriction is comparable with the Fermi wavelength $\lambda_F=\frac{2\pi\hbar}{\sqrt{2mE_F}}$ \cite{vanvwees}. In addition, the length, $L$, of the channel between the reservoirs must be smaller than or comparable to the  elastic mean free path $l_e$  for ballistic transport to take place. QPCs are nowadays routinely fabricated in semiconducting heterostructures such as AlGaAs/GaAs ones in which the very low carrier density ($\sim$10$^{11}$ cm$^{-2}$) results in large values of $\lambda_F$ ($\simeq$ 50 nm) \cite{rossler}. The extreme cleanliness of these heterostructures also results in a large mean free path that can exceed several micrometers at low temperatures \cite{rossler}. In contrast, the 2-DEG at the \LAO/\STO interface involves a higher carrier density  ($\sim$10$^{13}$ cm$^{-2}$) and has a reduced $l_e$ ($\sim$ 10-50 nm) imposing stronger constraints on the practical realization of such devices.  In this system, the conduction band is formed by coupling of the t$_{2g}$ orbitals (d$_\mathrm{xy}$, d$_\mathrm{xz}$ and d$_\mathrm{yz}$ orbitals) at neighbouring Ti lattice sites through the 2p orbitals of the oxygen atoms. Under strong quantum confinement in the direction perpendicular to the interface, the degeneracy of the t$_\mathrm{2g}$ bands is lifted, resulting in a rich and complex band structure with discrete 2D states separated by typical energies of tens of meV  \cite{salluzzo,berner}. Whilst multiband transport occurs at high carrier doping, in the depleted regime, only the lowest d$_\mathrm{xy}$ energy bands are filled. Figure 1 shows a schematic view of the QPC which is formed by applying a negative voltage on the split-gate to electrostatically deplete the 2-DEG underneath.  We used the  \LAO amorphous template method to fabricate a 10 $\mu$m wide Hall bar in a 12 u.c.-thick \LAO layer grown by Pulsed Laser Deposition on a  \TiO-terminated (001)-oriented \STO single crystal \cite{stornaiuolo, hurand}.  After the growth, a metallic back gate was deposited on the back side of the 500 $\mu m$ thick \STO substrate enabling a global control of the electron density in the device. Finally, a metallic split-gate was patterned by lift-off directly on top of the Hall bar (Fig. 1b).  Despite the reduced thickness of the \LAO layer ($\simeq$ 4.8 nm) and the absence of additional insulating dielectric layer, no significant leakage current was observed in this device ($\ll$1nA).  The separation between the two fingers at the centre of the split-gate is $W$ = 25 nm, which is comparable to the Fermi wavelength of the 2-DEG. Near the bottleneck of the constriction, the split-gate imposes a smoothly varying confining potential that can be modeled by a saddle form (Fig. 1d) \cite{buttiker}
 
\begin{equation}
V(x,y)=V_0-\frac{1}{2}m\omega_x^2x^2+\frac{1}{2}m\omega_y^2y^2
\label{potential}
\end{equation}
where $V_0$ is the potential at the centre of the QPC, $m$ is the electron mass and the frequencies $\omega_x$ and $\omega_y$ define the curvatures of the potential in the two directions. Because the potential along the $y$ direction is that of an harmonic oscillator, the lowest energy band of the 2-DEG is split into several 1D subbands. The number of conduction modes involved in the transport, which is related to the width of the constriction, is controlled by the split-gate voltage $V_\mathrm{SG}$.  For a given value of $V_\mathrm{SG}$,  channels corresponding to subbands of energies   $E_n=V_0+(n+1/2)\hbar\omega_y$, which are smaller than $E_F$, are open and participate to the conduction through the constriction.  In the framework of the Landauer-Buttiker formalism, the contribution of each channel to the conductance of the QPC is given by its transmission $T_n$ \cite{buttiker}:
\begin{equation}
T_n(E)=\frac{2}{1+e^{-\pi\epsilon_n}}
\label{quantization}
\end{equation}
where $\epsilon_n=2(E-E_n)/\hbar\omega_x$. 
The total conductance is the sum of the conductance of each of the channels,  $G=\sum_nT_nG_0$, where $G_0=\frac{2e^2}{h}$ is the quantum of conductance for spin degenerate subbands.\\

After fabrication, the sample was anchored to the mixing chamber of a dilution refrigerator and connected to coaxial lines that include low pass LC filters and copper powder filters at the lowest temperature stage. The differential conductance, $G(V_\mathrm{sd})$=$dI/dV_\mathrm{sd}$, was measured with a small ac excitation current using standard lock-in techniques in a four-terminal configuration. After cooling the sample, the back-gate voltage was first swept to a maximum value ($V_\mathrm{BG}$ = 15 V) to ensure that no hysteresis would take place upon further gating \cite{biscaras3}. During this forming step, the 2-DEG and the split gate are both kept at the ground ($V_\mathrm{SG}= 0$ V). The  back-gate voltage was then reduced to 8.1 V to strongly deplete the 2-DEG, and  the differential conductance, $G$, of the nanodevice was measured as a function of the dc source-drain voltage $V_{sd}$ and split-gate voltage $V_\mathrm{SG}$. For all the data presented in the following, the resistance of the reservoirs measured with the channel fully open ($V_\mathrm{SG} = 0.5$ V) was subtracted to obtain the intrinsic conductance of the QPC.  Figure 2  shows the evolution of the conductance at $V_\mathrm{sd}$ = 0 V for four different temperatures.   At $V_\mathrm{SG}$ = -0.2 V the QPC is pinched off. Plateaus corresponding to the quantized values of the conductance in steps of $G_0 =2e^2/h$ appear when the split-gate voltage is increased, which indicates that ballistic transport involving spin-degenerate bands is taking place in the QPC. As seen from the fit in Fig. 2 inset, this behaviour is well described by  the Landauer-Buttiker formalism (Eq. (2)) \cite{buttiker}. Oscillations of the conductance can be observed on the plateaus, which are likely related to interference and resonance effects associated with the geometry of the constriction \cite{kirczenow}.  An increase in temperature modifies the Fermi distribution and consequently the occupation of electronic states at the Fermi energy. This leads to a thermal broadening of the conductance steps between plateaus and, in addition, to a suppression of oscillations with temperature. The maximum number of conduction modes is related to the Fermi wavelength, $n_\mathrm{max}\simeq\frac{2W}{\lambda_F}$. As seen in Figure 2,  a maximum of three plateaus can be identified in this gate range, corresponding to $\lambda_F$ $\simeq$15 nm.  A clear quantization of the conductance, such as that observed here, only occurs  when $\omega_y>\omega_x$. This is confirmed by the fit shown in Fig. 2 inset, which provides a ratio of $\omega_y/\omega_x$ $\simeq$ 2.5. Previous studies have reported the fabrication of split-gate devices in \LAO/\STO but without achieving the quantization regime \cite{thierschmann}. \\

\begin{figure*}[tb]
  \includegraphics[width=\textwidth]{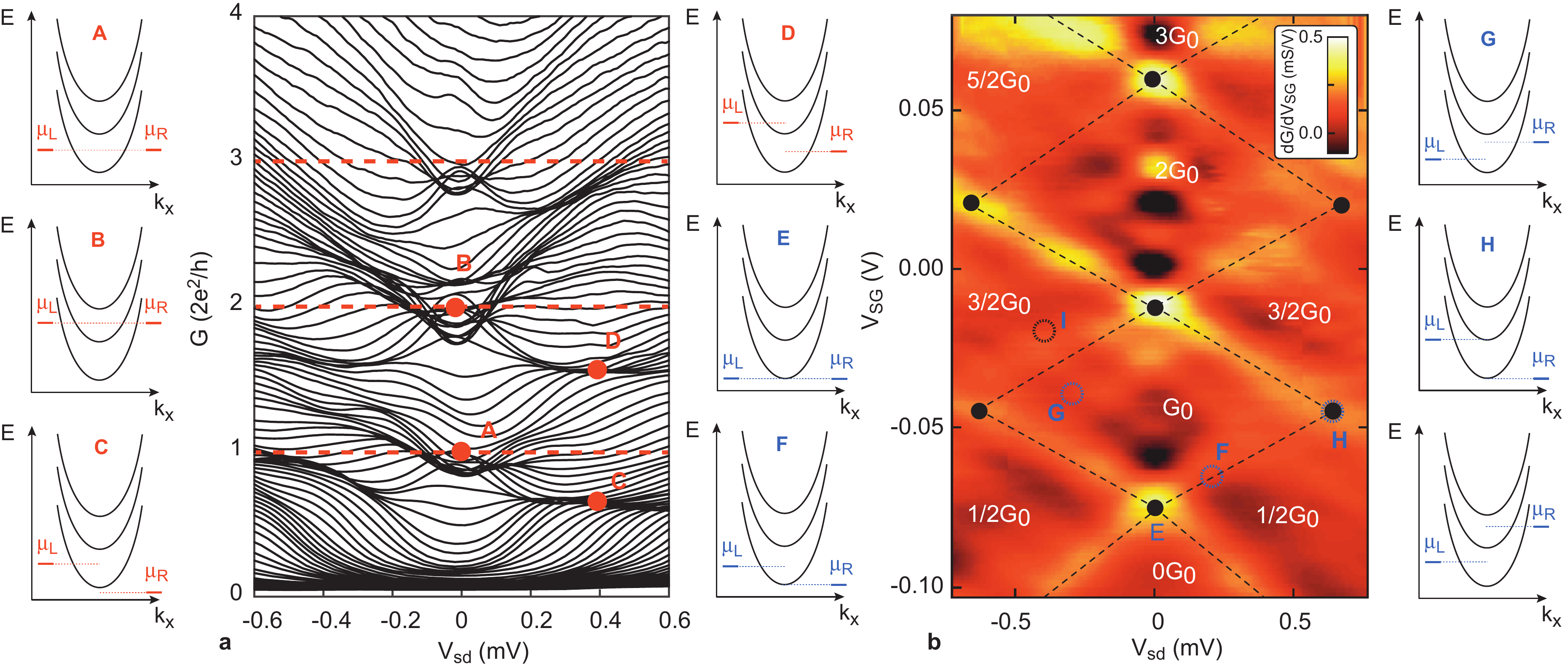}
  \caption{\textbf{Spectroscopy of the QPC subbands} a) Differential conductance of the QPC as a function of  source-drain voltage  for many values of the split-gate voltage in the range [-0.2 V, 0.1 V] and separated by an interval of 3 mV. Energy diagrams indicate the position of the chemical potentials in the left and right reservoirs corresponding to the different points labelled A,B,C and D. Integer  (e. g. points A and B) and half-integer  (e.g. points C and D) quantized conductance plateaus correspond to dark regions where curves accumulate. b)  Transconductance $dG/dV_\mathrm{SG}$ of the QPC  (color scale) as a function  of source-drain voltage and split-gate voltage. Dashed lines (guide for the eyes) highlight a diamond-like structure formed by joining light-colored points corresponding to maximum values of $dG/dV_\mathrm{SG}$. Diamonds centered around $V_\mathrm{sd}$ =0 V represent integer conductance plateaus  whereas external half-diamonds represent half-integer conductance plateaus. Energy diagrams show the position of the chemical potentials in the left and right reservoirs corresponding to the points labelled E, F, G, H and I on the color plot.}
\end{figure*}

The value of $\omega_y$ characterizing the harmonic confinement potential can be determined experimentally by measuring the conductance at non-zero source-drain voltage \cite{kouwenhoven,patel}.  In Fig. 3a, we plot a series of conductance curves, $G(V_\mathrm{sd})$ recorded for sequential steps in the split-gate voltage \cite{patel}.  Large nonlinearities are observed for applied biases comparable to the subband energy spacing. For $V_\mathrm{sd} = 0$ V, the  chemical potentials in the source and drain, $\mu_s$ and $\mu_d$ respectively, are aligned and we observe an accumulation of curves corresponding to the plateaus at integer values of $G_0$ (e. g. points ``A" and ``B" in Fig. 3a). At finite source-drain voltage, the chemical potentials of the left and right reservoirs are shifted by an amount $eV_{sd}$. When this energy exceeds the splitting between two subbands ($eV_\mathrm{sd} >\hbar\omega_y $), the number of transport modes available for left-going and right-going electrons differs by one, which produces additional half plateaus at conductance values of $(n+1/2)G_0$  \cite{kouwenhoven,patel,glazman}. This is seen at points labelled C and D in Fig. 3a, further confirming the quantization of the conductance.  Although these additional plateaus are predicted to be midway between those given by Eq. (2), some deviations of the conductance from half-integer values are also expected in particular for small numbers of channels \cite{glazman}. An asymmetry is observed in $V_{sd}$ which is likely due to an intrinsic asymmetry in the electrostatic potential defining the constriction.
\indent Non-linear transport can also be analysed by plotting the transconductance of the QPC, $dG/dV_\mathrm{SG}(V_\mathrm{sd})$, corresponding to the derivative of $G$ with respect to the split-gate voltage (Fig. 3b). In contrast to the series of conductance curves of Fig. 3a, which highlight the conductance plateaus at multiple values of $G_0$, the transconductance emphasizes the transitions between the plateaus. A set of regularly spaced points (in yellow), indicating a maximum in $dG/dV_\mathrm{SG}$, are connected by lines of high transconductance to form a diamond-like structure centered on $V_\mathrm{sd}$= 0 V (see dashed line). Each diamond represents a well defined quantized plateau where the chemical potentials of the two reservoirs lie in the same subband. On the other hand, regions outside the diamonds describe the situation where  $\mu_s$ and $\mu_d$ are aligned with different bands leading to conductance plateaus at $(n+1/2)G_0$. The crossing points (yellow) at  $V_\mathrm{sd}$ $\simeq$ 0 V corresponds to split-gate voltages at which the bottom of a new subband is aligned with the source and drain chemical potentials (e.g. point ``E"). The separation between two consecutive energy levels can be directly measured through  the source drain voltage at the edges of the diamonds. At point ``H" in Fig.3b, we read  $\hbar\omega_y$ = $eV_\mathrm{sd}(``H")\simeq$ 0.65 meV, which is consistent with the splitting between the d$_{xy}$ subbands due to lateral confinement in the QPC. Considering the ratio $\omega_y/\omega_x$ $\simeq$ 2.5 previously extracted from the fit in Fig. 2 inset, we obtain $\hbar\omega_x$ $\simeq$ 0.26 meV, which parametrizes the confinement potential of Eq. (\ref{potential}).

 We now investigate the effect of a magnetic field $B$ perpendicular to the 2-DEG plane, on the energy levels of the QPC. As seen in Figure 4a, the plateaus of conductance in the $G(V_\mathrm{SG})$ curve are now quantized in half-integer values of $G_0$, which indicates that the spin degeneracy is lifted. Consequently, the diamond-like structure is duplicated in the transconductance map  of Fig. 4c.  The Zeeman splitting between spin-polarized subbands can be directly measured by following the evolution of  transconductance at $V_\mathrm{sd}$ = 0 V as a function of the magnetic filed (Fig. 4a inset). The two spin bands are separated by an energy $\Delta E_z=g^*\mu_BB$, where  $\mu_B$ is the Bohr magneton and $g^*$ is the Land\'e g-factor, which describes the susceptibility of the electronic spin states to the external field, and which fundamentally governs spin-transport characteristics, e.g. spin coherence \cite{salis}.  Note that because of the reduced electron mobility in this depleted regime ($\mu\simeq$ 100 cm$^2$V$^{-1}$s$^{-1}$), the magnetic field used in this study is too weak to generate Landau levels and form hybrid magneto-electric subbands \cite{annadi}. 
 
  To access $\Delta E_z$ and calculate the g-factor, we measure $\delta V_\mathrm{SG}$  at 6 T (Fig. 4a inset) and convert it into an energy scale through the measurement of the splitting generated by the source drain bias $\frac{dV_\mathrm{SG}}{dV_\mathrm{sd}}$ as reported in Fig. 3b \cite{martin} . 
\begin{equation}
|g|=\frac{1}{\mu_B}\frac{\delta(\Delta E_z)}{\delta B}=\frac{e}{\mu_B}\frac{\delta V_\mathrm{sd}}{\delta V_\mathrm{SG}}\frac{\delta V_\mathrm{SG}}{\delta B}
\label{quantization}
\end{equation}

\begin{figure*}
  \includegraphics[width=13cm]{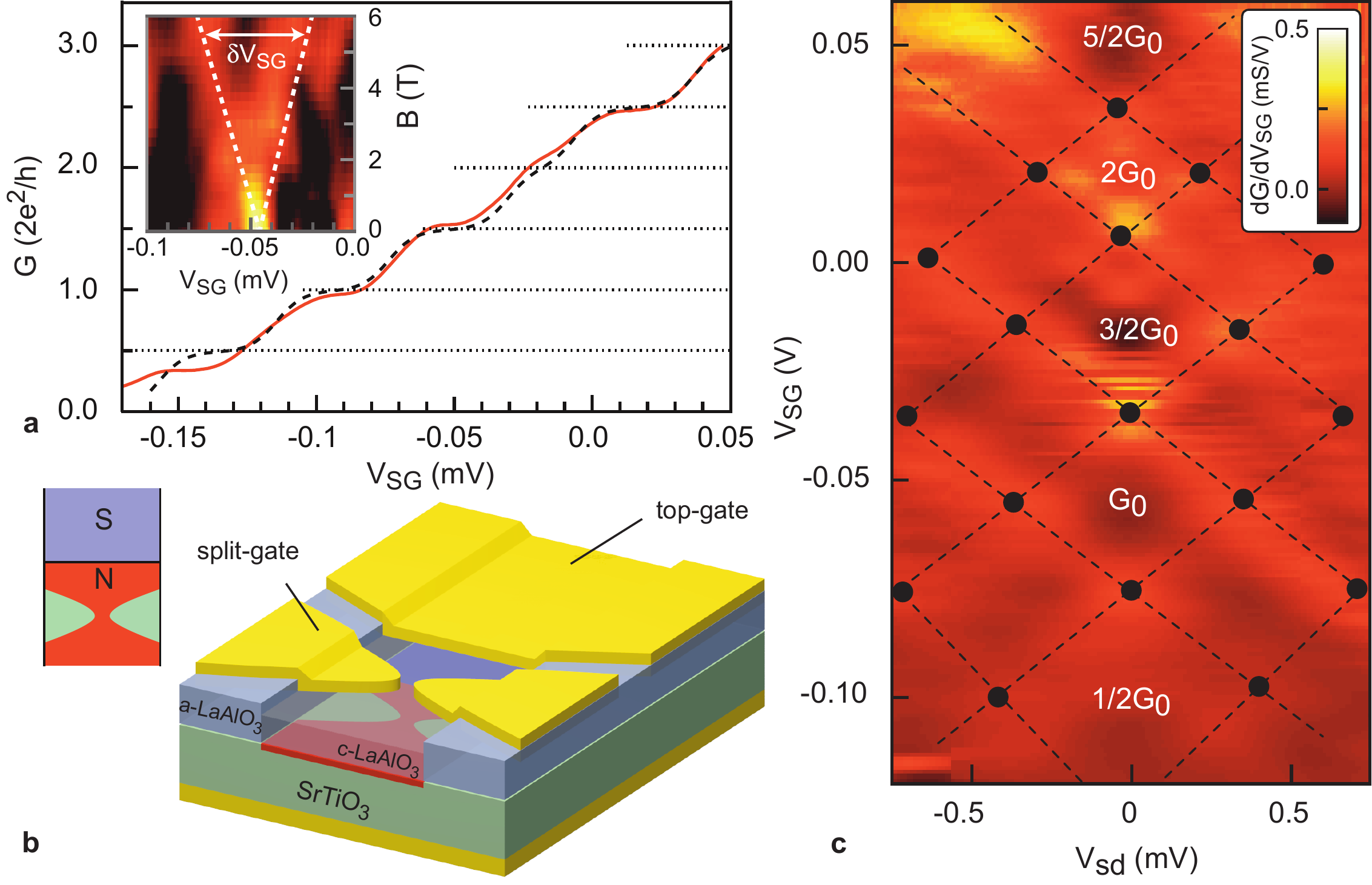}
  \caption{\textbf{Zeeman splitting and spin-polarized subbands under magnetic magnetic field}. a) Conductance of the QPC at $V_\mathrm{sd}$ = 0 V as a function of split-gate voltage $V_\mathrm{SG}$ for B = 6 T (red line) fitted by Eq. (2) with an additional Zeeman term (dashed line). Inset) Transconductance  $dG/dV_\mathrm{SG}$ of the $n$ = 2 step at $V_\mathrm{sd}$ = 0 V as a function of split-gate voltage and magnetic field.   b) Prototype device for the detection of Majorana states  in \LAO/\STO interfaces with a QPC. An additional top-gate is used to create, close to the QPC, a  junction between a superconducting region and a normal one with strong spin-orbit interaction. Inset) Scheme for detection of Majorana states  (taken from \cite{wimmer}). c) Transconductance $dG/dV_\mathrm{SG}$ of the QPC  (color scale) under a  B= 6T magnetic field as a function of source-drain voltage and split-gate voltage. The diamond-like structure formed by joining light-colored points corresponding to maximum values of $dG/dV_\mathrm{SG}$. Diamonds centered around $V_\mathrm{sd}$ =0 V represent conductance plateaus  in half integer values of $G_0$.}
  \label{fig3}
\end{figure*}

For the $n$ = 2 conductance step we find  $g^*$= 0.85 $\pm$ 0.1.  An estimate of the g-factor can also be obtained by fitting the conductance quantization curve of Fig. 4a, provided a Zeeman splitting term is introduced into Eq. (2). With this method, we obtain  a value $g^*$ = 0.9 $\pm$ 0.1, similar to that determined by the first method.  It is consistent with values obtained by non-direct methods reported in the literature.  For instance, in the low-doping regime, a g-factor between 0.5 and 1 was extracted from the analysis of the magneto-conductance in a weak-localization approach \cite{caviglia2}. A g-factor of 0.95,  close to our value, was also used to reproduce the experimental temperature dependence of  ARPES spectrum \cite{cancellieri}. The reduction of the g-factor with respect to the free electrons one is not currently understood in at \LAO/\STO interfaces but has often be attributed to the effect of spin-orbit coupling. Its low value could explain the high-value of the parallel superconducting  critical  magnetic field which has been found to exceed the Pauli limit if considering a conventional value of g $\simeq$2 \cite{kim,gervasi}. \\

In summary, we have demonstrated the realization of a QPC in the 2-DEG confined at the \LAO/\STO interface, which exhibits a clear conductance quantization due to  ballistic transport in a restricted number of conducting channels.  The characteristics energy scales of the confinement potential extracted from non-linear transport are consistent with a $d_{xy}$ band transport.  The measurement of the g-factor under a magnetic field performed on the $n$ = 2 conductance step provides a value in the range 0.85-0.9 for the g-factor, which differs from that of free electrons. The conductance curve with plateaus at multiples of $e^2/h$ shown in Fig. 4a, is characteristic of the normal state of a QPC under a magnetic field.  The demonstration of quantized conductance in oxide QPC  is a first step towards experiments to study topological superconductivity in oxide 2-DEGs.  Wimmer \textit{et al.} showed that when the QPC is put in close contact with a non-trivial topological superconducting region, e. g. an hybrid superconducting/semiconducting nanowire with large SOC or a superconductor characterized by large SOC (Fig. 4b inset),  the conductance plateaus should appear at $(2n+1)2e^2/h$ values \cite{wimmer,beenakker}.  In this case, the current flows through a single Majorana state localized in the vicinity of the QPC. The $n$ = 0 half integer plateau at a conductance of $2e^2/h$ is topologically protected against disorder which is expected to provide a robust distinctive signature of the topological phase. 
Since \LAO/\STO  is a superconductor with large SOC, such situation can be practically realized by an additional top-gate which triggers superconductivity in the oxide 2-DEG in close proximity to the QPC as proposed in the device of Fig. 4b. In addition, QPC devices in materials with strong SOC are also of strong interest to realize a spin-polarizer. In particular, spin separation of the electrons could be achieved in an asymmetric QPC quantum well in the presence of lateral SOC resulting in an electrically-controlled spin-sensitive device \cite{debray,ngo,kohda}.

\textbf{Acknowledgments}\\
The authors acknowledge M. Aprili, A. Caviglia, A. Akhmerov, R. Citro, M. Grilli, S. Caprara and L. Benfatto  for stimulating discussions. This work was supported by the French RENATECH network (French national nanofabrication platform), the R\'egion Ile-de-France in the framework of CNano IdF, OXYMORE and Sesame programs, by CNRS through a PICS program and by the ANR JCJC (Nano-SO2DEG). The Authors acknowledge received funding from the project Quantox of QuantERA ERA-NET Cofund in Quantum Technologies (Grant Agreement N. 731473) implemented within the European Union's Horizon 2020 Program. The authors also acknowledge the COST project Nanoscale coherent hybrid devices for superconducting quantum technologies-Action CA16218.\\

\end{document}